# A method using deep learning to discover new predictors of CRT response from mechanical dyssynchrony on gated SPECT MPI

Zhuo He, BS[a], Xinwei Zhang, MD, PhD[b], Chen Zhao, MS[a], Zhiyong Qian, MD, PhD[b], Yao Wang, MD, PhD[b], Xiaofeng Hou, *MD[b]*, Jiangang Zou, MD, PhD, FHRS[b*], *Weihua Zhou, PhD [a,c*]*

## *Institutions*

[a] College of Computing, Michigan Technological University, Houghton, MI

[b] Department of Cardiology, The First Affiliated Hospital of Nanjing Medical University, Nanjing, Jiangsu, China

[c] Center of Biocomputing and Digital Health, Institute of Computing and Cybersystems, and Health Research Institute, Michigan Technological University, Houghton, MI, 49931, USA

*Running Title:* Discover LVMD parameters for CRT

Zhuo He and Xinwei Zhang contributed equally to this work.

## *Address for correspondence*

Weihua Zhou

E-mail: whzhou@mtu.edu

Tel: 1-228-214-3265.

Address: 730 Beach Blvd E, Long Beach, MS 39560

or

Jiangang Zou

E-mail: jgzou@njmu.edu.cn

Tel. +86-13605191407.

Address: Guangzhou Road 300, Nanjing, Jiangsu, China 210029

# ABSTRACT

*Background.* Studies have shown that the conventional left ventricular mechanical dyssynchrony (LVMD) parameters have their own statistical limitations. The purpose of this study is to extract new LVMD parameters from the phase analysis of gated SPECT MPI by deep learning to help CRT patient selection.

*Methods.* One hundred and three patients who underwent rest gated SPECT MPI were enrolled in this study. CRT response was defined as a decrease in left ventricular end-systolic volume (LVESV) ≥ 15% at 6 ± 1 month follow up. Autoencoder (AE), an unsupervised deep learning method, was trained by the raw LV systolic phase polar maps to extract new LVMD parameters, called AE-based LVMD parameters. Correlation analysis was used to explain the relationships between new parameters with conventional LVMD parameters. Univariate and multivariate analyses were used to establish a multivariate model for predicting CRT response.

*Results.* Complete data were obtained in 102 patients, 44.1% of them were classified as CRT responders. AE-based LVMD parameter was significant in the univariate (OR 1.24, 95% CI 1.07 – 1.44, P = 0.006) and multivariate analyses (OR 1.03, 95% CI 1.01 – 1.06, P = 0.006). Moreover, it had incremental value over PSD (AUC 0.72 vs. 0.63, LH 8.06, P = 0.005) and PBW (AUC 0.72 vs. 0.64, LH 7.87, P = 0.005), combined with significant clinic characteristics, including LVEF and gender.

*Conclusions.* The new LVMD parameters extracted by autoencoder from the baseline gated SPECT MPI has the potential to improve the prediction of CRT response.

# ABBREVIATIONS

| | |
|---|---|
| AE | Autoencoder |
| CRT | Cardiac resynchronization therapy |
| ECTb4 | Emory Cardiac Toolbox Version 4.0 |
| LV | Left ventricular |
| LVEDV | Left ventricular end diastolic volume |
| LVEF | Left ventricular ejection fraction |
| LVESV | Left ventricular end systolic volume |
| PSD | Phase histogram standard deviation |
| PBW | Phase histogram bandwidth |
| MPI | Myocardial perfusion imaging |

# INTRODUCTION

Current guidelines recommend cardiac resynchronization therapy (CRT) for patients with low left ventricular ejection fraction (LVEF)(typically ⩽ 35%), sinus rhythm, left bundle-branch block (LBBB) with a QRS duration greater than or equal to 150 ms, and New York Heart Association (NYHA) class II, III, or ambulatory IV, symptoms on guideline-directed medical therapy.[1,2] However, 30-40% of the selected patients fail to respond.[3–5] In addition to symptoms of physical activity, left ventricular (LV) function, and electrical dyssynchrony measured by electrocardiogram (ECG), several studies have shown that left ventricular mechanical dyssynchrony (LVMD) of gated SPECT MPI has been well established in the prediction of CRT response. Henneman et al. demonstrated that baseline LVMD could improve the prediction of CRT response by a cutoff value of 43º for phase standard deviation (PSD) and 135º for phase bandwidth (PBW).[6] Many studies have further confimred that baseline PSD and PBW have predictive and prognostic values for CRT response.[5,7–10] The feasibility and effectiveness of LVMD in guiding LV lead implantation to improve CRT response have been validated by a large prospective multicenter trial.[11]

However, it was reported in the other prospective multicenter trial that baseline PSD and PBW were not associated with the CRT response, which defined by any improvement in one or more of the following: ≥ 1 NYHA class, increase in LVEF ≥ 5%, decrease in LVESV ≥ 15%, and ≥ 5 points in Minnesota Living With Heart Failure Questionnaire.[12] Gendre et al.[13] demonstrated that baseline LVMD parameter could not predict the response to CRT defined by a reduction of LVESV ≥ 15% or improve peak $VO_2 \geq 10\%$, even though the study is a single-center study with a small number of patients (n = 42). Zhang et al.[14] also found similar results in a multicenter study of 79 CRT patients, in which a reduction of LVESV ≥ 15% was used to define the volumetric response to CRT. The effectiveness of simple and conventional statistical methods to measure LVMD is still worth discussing.

Machine learning helps computers learn and develop rules without having to be instructed every step of the way by human programmers.[15] Deep learning combined many linear and non-linear transformations to obtain a more abstract and useful representation of data,[16] which is a powerful new machine learning tool with breakthrough application in disease detection and classification by absorbing the image measurement engineering directly into a learning step while processing the data in its natural form.[17–19] Betancur et al. present the effectiveness of using deep neural networks for feature extraction in gated SPECT MPI and predict obstructive CAD.[20] Xu et al. demonstrated that unsupervised feature extraction by deep learning is as effective as a supervised method.[21] To the best of our knowledge, there has been no work that uses deep features for CRT patient selection by gated SPECT MPI data. In this study, we aimed to extract new phase parameters from phases analysis of gated SPECT MPI by deep learning and explore the potential of extracted new phase parameters to predict CRT response.

# METHODS

## Patient population

One hundred and four CRT patients with gated resting SPECT MPI were enrolled at the First Affiliated Nanjing Medical University Hospital from July 2008 to July 2017. The enrolled criteria including (1) Left ventricular ejection fraction (LVEF) measured by echocardiography ≤ 35% (2) QRS duration ≥ 120 ms; (3) New York Heart Association (NYHA) functional class from II to IV; and (4) optimal medical therapy for at least 3 months before CRT implantation. Patients with atrial fibrillation, right bundle branch block were excluded.

All the patients had baseline data of clinical characteristics, echocardiography, resting gated SPECT MPI and 6 months follow-up data of clinical characteristics and echocardiography. The study was approved by the Institutional Ethical Committee of the First Affiliated Hospital of Nanjing Medical University, and informed consent was obtained from all patients.

## Evaluation of LV function by echocardiography

Echocardiography data of all patients were assessed by experienced ultrasound experts blinded to any clinical data and MPI data before and 6 months after CRT. LV end-diastolic volume (LVEDV), LV end-systolic volume (LVESV), and LV ejection fractions (LVEF) were measured by the 2-dimensional modified biplane Simpson method. In 6-month follow-up echocardiography, a decrease of ESV ≥ 15% was considered a positive mechanical response to CRT, and it has been widely accepted as the mechanical response for CRT. [5,22,23]

## Gated SPECT MPI acquisition and phase dyssynchrony analysis

Gated SPECT MPI was performed by conventional SPECT systems with a low-energy, general-purpose collimator within seven days before CRT implantation. Resting gated SPECT scan was performed 60-90 minutes after injection of 25-30mCi of 99mTc-MIBI, which were acquired by one-day resting gated SPECT MPI protocol with a dual-headed or triple-head camera by 180° orbits with a complimentary 8 frames ECG-gating, according to the current guideline.[24] MPI were acquired by a photopeak window of the 99mTc set as a 20% around 140KeV. All the images were reconstructed by the OSEM method from Emory

Reconstruction Toolbox (ERTb2, Atlanta, GA) with 3 iterations, 10 subsets, power 10, and a cutoff frequency of 0.3 cycles/mm. Reorientation into short-axis images was processed by Emory Cardiac Toolbox (ECTb4, Atlanta, GA) for automated measurement of perfusion, function, and phase dyssynchrony analysis.

Phase dyssynchrony analysis includes three steps: 1) extracting three-dimensional maximal count myocardial perfusion from the short axis slices by the sampling of the myocardial wall; 2) calculating the systolic or diastolic dyssynchrony phase angles by the 1- or 3-harmonic Fourier approximation, respectively, which measures the change of counts in LV myocardium in an R-R cardiac cycle; 3) generating polar maps of LV phase based on the 3D phase angle distribution in the LV myocardium.[25,26]

**Deep learning model**

The overall process of the deep learning model is illustrated in Figure 1. Extraction of LVMD parameters was accomplished by autoencoder (AE), which is a type of unsupervised neural network used to learn efficient image features by copying the inputs to the outputs and compressing the input into a latent-space representation, and then reconstructing the output from this representation.[27]

The network was arranged in two stages, the encoder and the decoder. The encoder connects directly with the phase polar map pixels via linear layer, compresses the high dimensional input images into a latent variable embedding, which has a lower dimension than the input images and is used as a characterization of the input images. The decoder attempts to reconstruct the input images from the produced features. The AE model learned to recognize key polar map features and reconstruct the input by minimizing the error between the reconstructed output and input images.

For extracting features, we use a four-layered linear autoencoder with an iteration set to 1000 as the parameters. To extract features, we gave all input to the autoencoder and trained our network for 104 instances (1 instance =1 phase polar map). Then we extracted features from hidden layers of the autoencoder. The features from layer 2 of the autoencoder of 4 layers were used to create a feature vector of 64 dimensions for each instance.

The AE model was implemented by the PyTorch deep learning toolkit (version 1.3.1) in the Python programming language.[28] Model training was performed on graphical processor units (TITAN Xp, NVIDIA, Santa Clara, California).

**Comparison of clinical and deep learning variables by statistical analysis**

Non-continuous variables of the baseline characteristics were expressed in number and percentage and tested with the Chi-square test. Continuous variables were expressed as mean ± standard deviation and tested with the Student t-test. After obtaining the significant AE-based LVMD parameters, the Pearson correlation coefficient was used to determine whether there were differences between them and the conventional LVMD parameters. The univariate logic regression analysis was applied to estimate the predictive values of baseline clinical variables, conventional LVMD parameters, and AE-based LVMD parameters for CRT response. Although the AE method has converted high-dimensional images into relatively low-dimensional features, the number of these AE-based features is still too high to compare them with current LVMD parameters. In order to overcome the curse of dimensionality, only significant features in univariate analysis were inputted into the prediction model of CRT response and analyzed in a clinical sense. Variables with $p < 0.05$ were considered statistically significant, and variables with $p < 0.1$ were included in the multivariate logic regression analysis. The predictive performances of all significant variables were evaluated by the area under the curve (AUC) of the receiver operating characteristic (ROC) of binary logistic regression. Statistical analysis was performed by Python Statsmodels package.[29]

# RESULTS

## Baseline characteristics

Among all 103 CRT patients, one patient had an extremely small ESV (16mL), which was considered as an outlier caused by the low resolution of gated SPECT MPI. The baseline characteristics of 102 patients were shown in Table 1. The average age was 61.3 ± 13.0 years, 74 (72.5%) patients were male, and 56 (54.9%) patients were classified as NYHA functional class III.

After 6-mont follow-up, 45 of the 102 patients (44.1%) were considered as CRT response, and 57 55.9%) were considered as non-responders to CRT. Significant differences of QRS duration (165.4 ± 16.2 vs. 150.9 ± 23.7, P = 0.001), ESV (270.5 ± 81.8 vs. 307.5 ± 90.0, P = 0.036) and EDV (191.5 ± 71.6 vs. 222.6 ± 77.7, P = 0.043) were noted between the two groups. However, there were no significant differences between the two groups in baseline PSD and PBW.

## Statistical analysis

In the univariate analysis, none of the variables were considered significant to CRT response, except for three AE-based LVMD parameters, which were shown in Table 2. Pearson's correlation coefficient has shown that the AE-based LVMD parameters were totally different from conventional LVMD parameters, including PSD and PBW. Due to the collinearity of AE parameters, AE # 30 was analyzed with other clinic variables alone, and AE 28 and AE #35 were combined and analyzed together with other clinic variables. Only AE #30 (OR 1.03, 95% CI 1.01 – 1.06, P = 0.006) was significant in multivariate analysis, as shown in Table 2.

In ROC analysis of AE-based LVMD parameters, the area under the curve (AUC) of the combination of clinical variables with conventional LVMD (PSD: AUC 0.63, P = 0.003; PBW: AUC 0.64, P = 0.003) were much lower than the combination of clinical variables with AE-based LVMD parameters (PSD: AUC 0.72, P = 0.005; PBW: AUC 0.72, P = 0.005). Moreover, sequential models indicated that the addition of AE-based LVMD parameter (AE #30) had incremental value over PSD (Likelyhood [LH] 8.06, P = 0.005), PBW (LH 7.87, P = 0.005) to predict CRT response, and the combination of AE #28 and AE # 35 also had incremental value over PSD (LH 9.60, P = 0.008) and PBW (LH 9.45, P = 0.009).

## DISCUSSION

In this research, we presented a CRT patient selection system for classifying patients by phase polar map as either CRT candidates or not. The proposed system uses deep learning to extract features from an autoencoder. We showed that the proposed system convincingly outperforms the state-of-art method.

## Limitations of LVMD

LVMD has different effects on predicting the response of CRT in different studies. Sillanmäki et al.[30] found that QRS duration was not independently associated with LVDM; even LVEF had shown a higher impact than QRS duration on LVMD. Despite the fact that early retrospective studies suggested a significant association between baseline LVMD and CRT response,[31] a study do not support this and instead suggest a strong relationship between QRS duration and CRT response.[32] Moreover, Bleeker et al. demonstrated that the correlation between QRS duration and LVMD was poor.[33] Therefore, it is seemingly disparate that select CRT patients base on conventional LVMD or QRS duration when there appears to be a poor correlation between them.

From the calculation method of LVMD, using statistical methods was easily affected by the outliers of phase measurement.[34–36] PSD may be deceptive for characterizing the widely distributed and multi-modal distributions in phase histograms; PSD, PBW, and phase entropy are significantly different in gender and related to LV volume and LVEF. The statistical analysis relies heavily on processing quantitative data, typically done by checking the relationship between factors or by grouping empirical data into categories.[37] There are problems in identifying appropriate data, processing interconnected rather than independent factors, and even violations of the assumption.

A method that comprehensively considers the relationship between variables and extracts potential, new and more predictive features for CRT is what we need now.

## Advantages of deep learning

Compared with conventional statistical methods, data-driven feature learning through deep learning has higher performance. The advantage of deep learning for feature learning is a layered architecture similar to the human brain.[38] Through deep learning, simple features are extracted from the raw data, and then more complex features are learned through multiple layers. Finally, considerable features are generated through multiple iterations of learning. Specifically, feature learning is classified into two categories, supervised learning and unsupervised learning.

In supervised learning, the data is forwarded from input to output for prediction. By minimizing the value of the cost function between the target value and the predicted value, back propagation is used to optimizes the parameters of the deep learning model. Betancur et al.[20] proposed a deep convolutional neural network for predicting the probability of obstructive coronary artery disease in the left anterior descending artery (LAD), left circumflex artery (LCx), and right coronary artery (RCA), which had better per-patient sensitivity (from 79.8% to 82.3%, $p<0.05$) and per-vessel sensitivity (from 64.4% to 69.8%, $p<0.01$). However, it is difficult to collect the target value, which is the correct answer labeled manually, for a large dataset.

In unsupervised learning, unlabeled data is used to learn features on its own, which can give us unknown information. Cikes et al.[39] proposed an unsupervised approach to analyze the phenotype of heterogeneous HF by integrating clinical characteristics and eight echocardiographic descriptors (traces) extracted from full heart cycle echocardiographic images. They emphasized that their unsupervised method was not trained based on a priori knowledge, but the interpretability of their model was based on the distribution of existing prior variables in different phenogroup.

This paper proposes a linear autoencoder unsupervised learning algorithm for CRT features learning and interprets these new features' clinical meaning by statistical models. The unsupervised deep learning method, autoencoder, is learned automatically from SPECT MPI. It means that it is easy to train specific instances of the algorithm that will perform well on a particular type of patient and that it only requires appropriate unlabeled training images. And the data-driven method is available to avoid the loss of important information in feature extraction rather than global conventional statistical parameters, which are too simple to demonstrate the contraction pattern. Moreover, Pearson's correlation, univariate and multivariate analysis proved that the newly extracted variables from the deep learning model are different from conventional phase parameters and have a significant relationship with CRT response. Furthermore, these LVMD variables have incremental value over conventional LVMD parameters.

**Limitations**

Several limitations of this study should be acknowledged. This study enrolled a relatively small number of patients from multicenter with the inherent limitation of such study design. The performance and feasibility of the data-driven system may be affected by the quality of data, and further validation in other data is needed to evaluate new features extracted from the deep learning model to predict CRT response. Moreover, because the role of experts is ignored, we will combine domain knowledge and data-driven feature learning in our feature work.

# CONCLUSION

The new LVMD parameters extracted by autoencoder from the baseline gated SPECT MPI has the potential to improve the prediction of CRT response.

# NEW KNOWLEDGE GAINED

The role of LVMD for CRT patient selection remains unclear. This paper proposed a novel approach to extract new LVMD variables from gated SPECT MPI by unsupervised autoencoder deep learning. These new variables are different from existed LVMD parameters and confirmed to be independent predictive variables for CRT response.

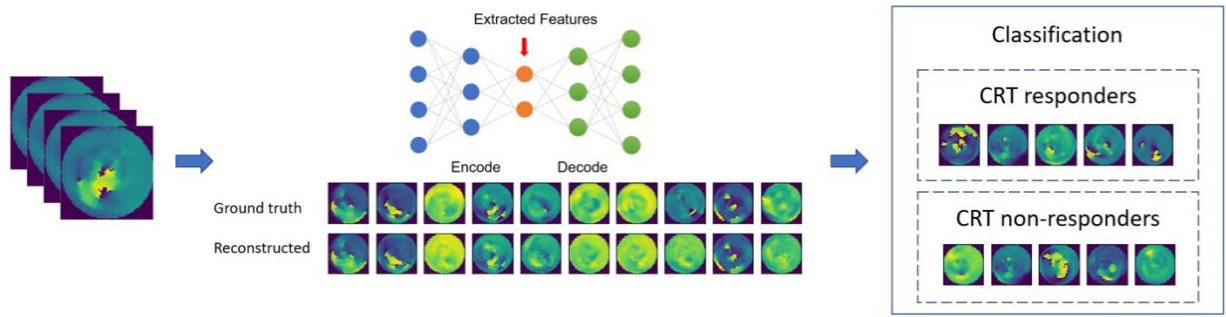

Figure 1. Block diagram of CRT response prediction model by the autoencoder

Table 1. Baseline characteristics and LV parameters of the enrolled patients.

| Variables | All(n=102) | Response (n=45, 44.1%) | Non-response (n=57, 55.9%) | P-value (T-test/Chi^2) |
|---|---|---|---|---|
| ACEI/ARB | 88 (86.3%) | 37 (82.2%) | 51 (89.5%) | 0.443 |
| Age | 61.3 ± 13.0 | 62.0 ± 12.4 | 60.8 ± 13.5 | 0.672 |
| CAD | 25 (24.5%) | 10 (22.2%) | 15 (26.3%) | 0.806 |
| CKD | 5 (4.9%) | 3 (6.7%) | 2 (3.5%) | 0.786 |
| QRS duration | 157.3 ± 22.0 | 165.4 ± 16.2 | 150.9 ± 23.7 | 0.001 |
| EDV | 291.2 ± 88.4 | 270.5 ± 81.8 | 307.5 ± 90.0 | 0.036 |
| ESV | 208.9 ± 76.6 | 191.5 ± 71.6 | 222.6 ± 77.7 | 0.043 |
| LVEF | 29.3 ± 6.8 | 29.9 ± 6.7 | 28.8 ± 6.8 | 0.409 |
| Gender | 74 (72.5%) | 37 (82.2%) | 37 (64.9%) | 0.085 |
| Ischemia | 24 (23.5%) | 9 (20.0%) | 15 (26.3%) | 0.609 |
| NYHA |  |  |  | 0.330 |
|   II | 31 (30.4%) | 16 (35.6%) | 15 (26.3%) |  |
|   III | 56 (54.9%) | 21 (46.7%) | 35 (61.4%) |  |
|   IV | 15 (14.7%) | 8 (17.8%) | 7 (12.3%) |  |
| Scar burden | 26.8 ± 11.2 | 24.5 ± 9.8 | 28.6 ± 11.9 | 0.067 |
| PBW | 195.4 ± 75.9 | 197.1 ± 72.8 | 194.1 ± 78.2 | 0.845 |
| PSD | 58.1 ± 18.7 | 58.6 ± 18.3 | 57.7 ± 19.0 | 0.819 |
| Smoking | 37 (36.3%) | 19 (42.2%) | 18 (31.6%) | 0.367 |

Table 2. Univariate and multivariate logistic regression analysis.

| Variables | Univariate analysis | | | Multivariate analysis (AE #30) | | | Multivariate analysis (AE #28 #35) | | |
|---|---|---|---|---|---|---|---|---|---|
| | OR | 95% CI | P value | OR | 95% CI | P value | OR | 95% CI | P value |
| **AE #28** | 1.24 | [1.03, 1.48] | 0.02 | | | | 1.03 | 1.00 – 1.07 | 0.059 |
| **AE #30** | 1.24 | [1.07, 1.44] | 0.006 | 1.03 | 1.01 – 1.06 | 0.006 | | | |
| **AE #35** | 1.2 | [1.01, 1.41] | 0.034 | | | | 1.03 | 1.00 – 1.05 | 0.052 |
| **Scar burden** | 0.99 | [0.95, 1.02] | 0.505 | | | | | | |
| **ACEI/ARB** | 1.63 | [0.52, 5.16] | 0.402 | | | | | | |
| **Age** | 1 | [0.97, 1.03] | 0.96 | | | | | | |
| **CAD** | 0.84 | [0.33, 2.16] | 0.715 | | | | | | |
| **EDV** | 1 | [0.99, 1.0] | 0.283 | | | | | | |
| **ESV** | 1 | [0.99, 1.0] | 0.622 | | | | | | |
| **LVEF** | 0.95 | [0.89, 1.01] | 0.091 | 0.99 | 0.98 – 1.00 | 0.358 | 0.99 | 0.98 – 1.01 | 0.307 |
| **Gender** | 2.23 | [0.91, 5.47] | 0.08 | 1.16 | 0.94 – 1.42 | 0.153 | 1.17 | 0.95 – 1.43 | 0.129 |
| **NYHA** | 1.01 | [0.55, 1.85] | 0.986 | | | | | | |
| **PBW** | 1 | [1.0, 1.01] | 0.112 | | | | | | |
| **PSD** | 1.02 | [1.0, 1.04] | 0.123 | | | | | | |
| **QRS duration** | 1.01 | [0.99, 1.03] | 0.157 | | | | | | |
| **Smoking** | 1.9 | [0.77, 4.67] | 0.164 | | | | | | |

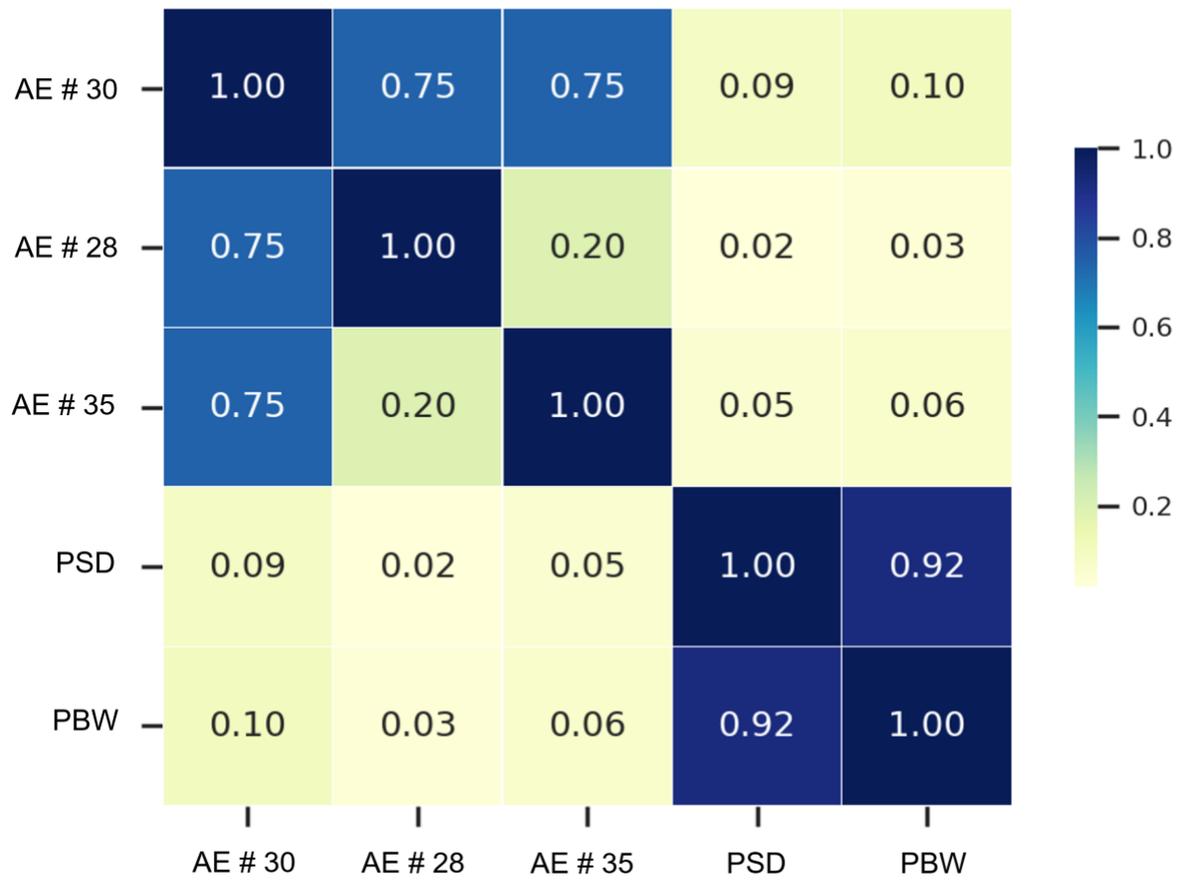

Figure 2. Pearson's correlation between PSD, PBW, and AE-based LVMD parameters

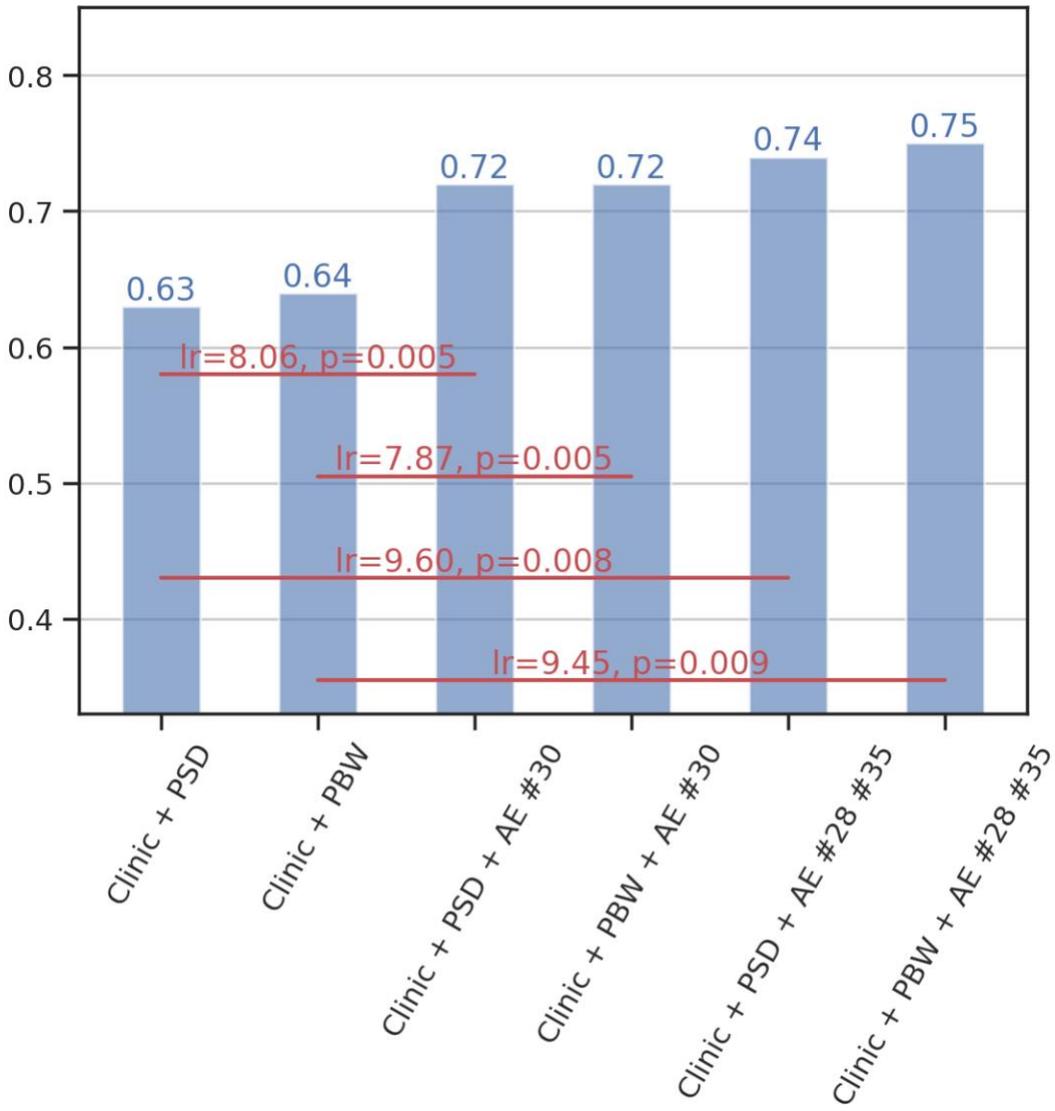

Figure 3. Incremental adjusted additive value of autoencoder phase parameters in the prediction of CRT response.